\documentclass[twocolumn,twocolappendix]{aastex631}
\usepackage{amsmath}

\newcommand\kms{km s$^{-1}$}

\newcommand\teff{$T_{\rm eff}$}
\newcommand\logg{$\log g$}

\newcommand\msun{M$_\odot$}
\newcommand\rsun{R$_\odot$}
\newcommand\av{$A_V$}

\newcommand\eba{2M05$-$06}
\newcommand\ebb{2M05$-$00}

\defcitealias{kounkel2018a}{Paper I}
\usepackage{listings}

\begin{document}
%\title{Two unusual young eclipsing binaries in Orion}
\title{Two young eclipsing binaries in Orion with temperatures and radii affected by spots and third bodies}

\author[0000-0002-5365-1267]{Marina Kounkel}
\affil{Department of Physics and Astronomy, University of North Florida, 1 UNF Dr, Jacksonville, FL, 32224, USA}
\email{marina.kounkel@unf.edu}
\author[0000-0002-3481-9052]{Keivan G.\ Stassun}
\affiliation{Department of Physics and Astronomy, Vanderbilt University, Nashville, TN 37235, USA}

\begin{abstract}
In this work we present a model of two young eclipsing binaries in the Orion Complex. Both heavily spotted, they present radii and temperatures that are in disagreement with the predictions of standard stellar models. \eba\ consists of two stars with different masses ($\sim$0.52 and $\sim$0.42~\msun) but with very similar radii ($\sim$0.9~\rsun), and with the less massive star having a highly spotted surface that causes it to have a hotter (unspotted) photosphere than the higher-mass star. The other system, \ebb, consists of two stars of very similar masses ($\sim$0.34~\msun), but very different radii ($\sim$0.7 and $\sim$1.0~\rsun), which creates an appearance of the two eclipsing stars being non-coeval. 
\ebb\ appears to have a tertiary companion that could offer an explanation for the unusual properties of the eclipsing stars, as has been seen in some other young triple systems.
Comparing the empirically measured properties of these eclipsing binaries to the predictions of stellar models, both standard and magnetic, we find that only the magnetic models correctly predict the observed relationship between mass and effective temperature. However, standard (non-magnetic) models better predict the temperatures of the unspotted photospheres. These observations represent an important step in improving our understanding of pre--main-sequence stellar evolution and the roles of spots and tertiaries on fundamental stellar properties. \\
\end{abstract}


\section{Introduction}

Mass and radius are some of the most fundamental properties of a star. In pre-main sequence (PMS) stars in particular, measuring these parameters is vital in order to understand their evolution. Obtaining accurate, unbiased, and model-independent measurements, however, is difficult to do for most young stars. Typically, these parameters are interpolated from using a combination of photometric or spectroscopic properties relative to the theoretical isochrones. But, these models of isochrones do not perfectly agree with each other and they often offer an imperfect fit to the data, e.g., due to prominent spots on the surface of PMS stars. 

Thus, a census of young stars with independently derived stellar properties vital to improve the calibration of these models. One of the most direct methods of measuring them is though analyzing the orbits of eclipsing double lined spectroscopic binaries (ESB2s): in them it is possible to solve for masses of the individual stars, their radii, their surface temperatures, the orbital separation, orbital period, inclination, and eccentricity, all directly from the data with no model assumptions. Unfortunately, at the current time the census of low mass PMS ESB2s is still quite small, amounting to only $\sim$30 systems, in large part to the great volume of photometric time series data that is needed to observe eclipses, combined with high resolution multi-epoch spectroscopy. A larger, more statistical census of young ESB2s is needed for a more comprehensive analysis.

In this paper we present a characterization of two such system, 2MASS J05351685$-$0618158 (hereafter \eba) and 2MASS J05454167$-$0004024 (hereafter \ebb), both associated with the Orion Complex. In Section~\ref{sec:data} we describe the light curves and radial velocities used in this work. In Section~\ref{sec:analysis} we describe the fitting procedure to determine the masses, radii, and other stellar and orbital parameters. In Section~\ref{sec:result} we present the parameters of both of these systems. In Section~\ref{sec:discussion} we discuss the implications of these parameters on the current evolutionary models, and present comparison with other young EBs. Finally, in Section~\ref{sec:concl} we conclude our resuts.

\section{Data}\label{sec:data}

\begin{figure*}
\epsscale{1.2}
\plotone{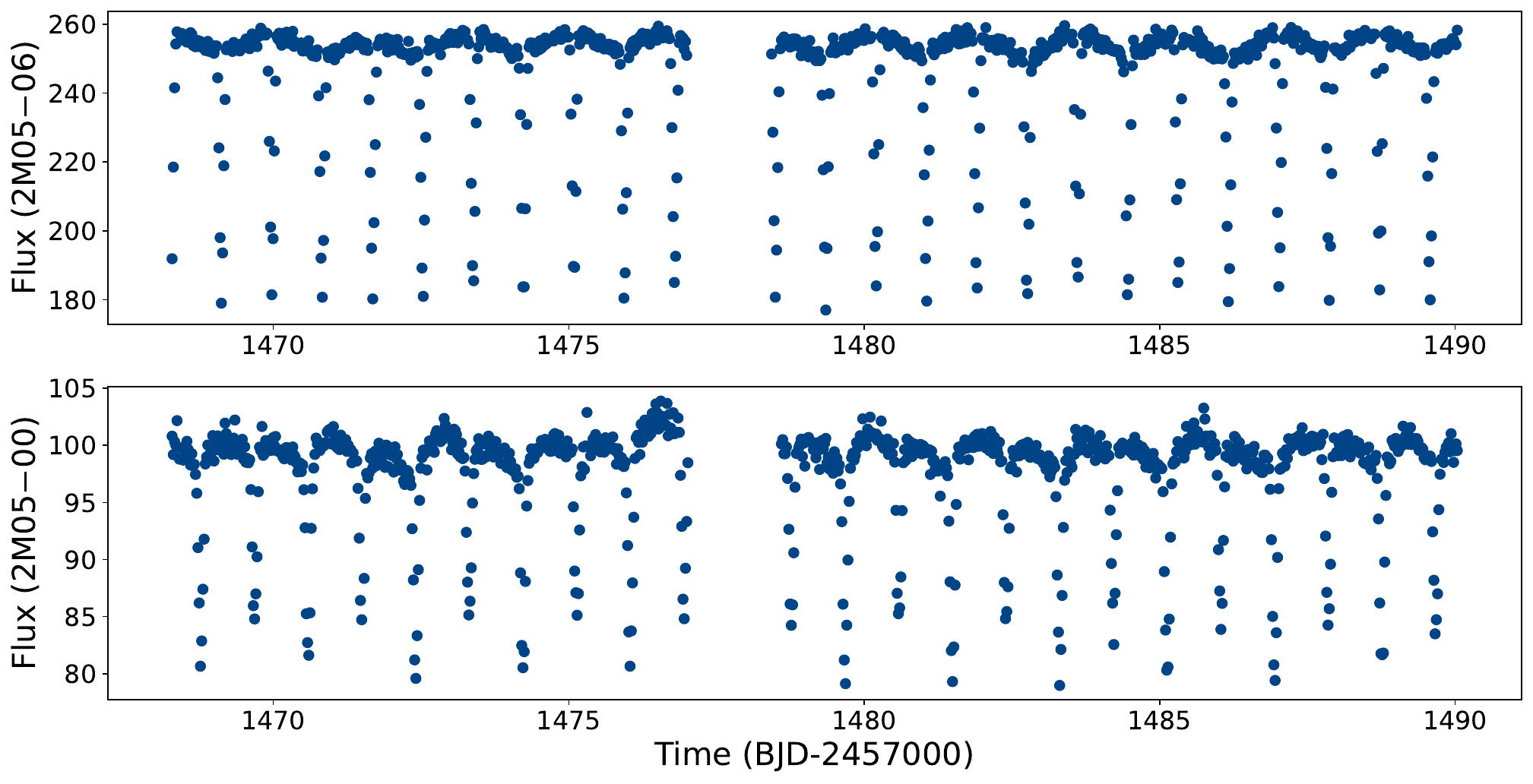}
\caption{TESS light curves for \eba\ (top) and \ebb\ (bottom)
\label{fig:tess}}
\end{figure*}

APOGEE is a high resolution spectrograph with $R\sim22,500$ covering the wavelength range of 1.5--1.7 $\mu$m that is capable of observing up to 300 sources simultaneously in a 2.5$^\circ$ field of view. It has performed a survey of the Orion Complex, obtaining multi-epoch spectra for $\sim$3500 young stars in this star forming region \citep{kounkel2018a}. In these data, \eba\ and \ebb\ have previously been identified as SB2s in \citep{kounkel2021}, with the full reported complement of the individual RVs for both primary and the secondary in both systems. A total of 7 and 6 epochs are available for both of them respectively; they can be resolved as SB2s in all of them. Through the course of the analysis we had to reassess which RVs were assigned to the primary and which to the secondary to fix mismatched assignments in several epochs in \citet{kounkel2021} for \eba.

Through spectral modeling of the APOGEE spectra, \eba\ has been reported to have \teff=3300$\pm$25 K and \logg=4.17$\pm$0.07. For \ebb, \teff=3400$\pm$30 K and \logg=4.14$\pm$0.03 \citep{sprague2022}; this is in great agreement with their spectral types derived from lower resolution optical spectra of M4 \citep{fang2009} and M3 \citep{hsu2012} respectively \citep[which convert to 3200 K and 3400 K][]{pecaut2013}.

Upon examining light curves from TESS Sector 6 full frame images constructed with eleanor \citep{feinstein2019}, these two systems showed the presence of eclipses, with the respective period of 1.7044 and 1.817138 days(Figure \ref{fig:tess}). Time series data is also available for them as part of Gaia DR3 \citep{gaia-collaboration2022a}; while they are not labelled as eclipsing in the data release, through being folded onto a correct period, eclipses in all 3 bandpasses (G, BP, RP) are also apparent (Figures \ref{fig:eb1},\ref{fig:eb2}). 

TESS light curves for both stars exhibit significant asymmetry in the continuum, indicative of the rotational variability with the timescale comparable to the orbital period. Despite their youth, rotation of both binaries appear to be tidally locked to their respective orbit. In Gaia passband, the effect is less apparent due to the sparcity of the data making it difficult to conclusively observe the shape of the continuum.

\section{Analysis}\label{sec:analysis}

\subsection{Basic Setup}

We used PHOEBE \citep{conroy2020} to perform the analysis of the light \& RV curves for each system. There are a total of 10 orbitals parameters we considered for each system. They include individual radii, $R_1$, $R_2$, their temperatures \teff$_{,1}$ \teff$_{,2}$, orbital period $P$, time of passage of the superior conjunction $t_0$, semi-major axis $a$, inclination $i$, mass ratio $q$, RV of the system $v_0$. Eccentricity $e$ and argument of periastron $\omega$ were initially considered, but were deemed unnecessary as the systems appear to have circular orbits. We also fit third light contamination $l3$ for each of the four light curves: in particular, since TESS has low spatial resolution of 21''/pixel, care needs to be taken by accounting for contamination in denser regions such as star clusters or associations. We also include passband luminosity for all 4 light curves, this is a nuisance parameter corresponding to the scaling of the continuum flux, which is recommended for stability within PHOEBE. 

Finally, since these systems do appear to exhibit rotational variability, in both systems we also fit spot parameters on each star: including spot longitude ($l$) and latitude ($b$), and spot radius ($r$). During the fitting the spot temperature contrast was fixed to 0.85, which is typical for spots in stars of these \teff\ \citep{berdyugina2005}. 

In what follows, we discuss the treatment of \teff\ and of spots in greater detail, and then summarize the final fitting approach.

\subsection{\teff\ from Spectral Energy Distribution Fitting}

Eclipsing binaries make it possible to precisely determine the ratio of \teff\ of the primary and the secondary through considering the differences between the eclipse depths, even when all that is available is just a light curve at a single bandpass \citep[e.g.,][]{prsa2008}. However, while it is also possible to determine ``true'' \teff\ of each individual star given sufficient data about the system (such as RVs, as well as light curves at different bandpasses), they are much more difficult to constrain than \teff\ ratio. In practice, absolute calibration of \teff\ often requires an independent fit of the spectral energy distribution (SED) to place stronger constraints on the limits of plausible \teff s for the sytem \citep[e.g.][]{kounkel2024}, as without it, the model may significantly underestimate or overestimate \teff s (while preserving \teff\ ratio). Unfortunately, however, PHOEBE does enable consideration of single epoch broadband photometry when modeling the system, requiring an independent SED fitting procedure, which may then require iteratively recomputing a model within PHOEBE with a stronger prior in order to converge.

We thus use SEDFit \citep{sedfit}. It downloads all of the available photometry from UV to mid-IR and considers a model photosphere using BT-Settl synthetic grid \citep{allard2011}. These models are interpolated to \teff\ and \logg\ of a star, scaled to the appropriate radius, fitting for the distance and extinction. In case of multiple stars, different templates are added together.

\subsection{Spots}

The situation is made more complex when spots are considered. PHOEBE does not report a ``true'' effective temperature in spotted stars, rather what it refers to as \teff\ is the \teff\ of the unspotted portion of the photosphere. But, since PHOEBE can model a spot independently, with the spot being of a given radius and given temperature contrast, we recalculate \teff\ in Table \ref{tab:params} with spots in mind across the entire surface of a star as $(\int T^4 dA / A)^{0.25}$ as per definition. And, for the purposes of SED fitting, to avoid biases, we compute a mesh for each star, each surface in the mesh having an independent estimate of temperature, \logg, and surface area; a synthetic model is computed for each surface, and the flux of each of them is added together for a given star.

We note that whether a spot is being fitted or not, the stellar parameters that are determined by PHOEBE are typically not affected; it generally produces an almost identical fit for masses, radii, temperature ratios, third light, and all of the other orbital parameters \citep{kounkel2024}, which we were able to confirm in the system in this work. Thus, through iterative exploration of the parameter space, considering spots of different sizes alongside with different PHOEBE models and the quality of fit of the resulting SED, we place the prior on the temperature of the photosphere to be 3000--3800 K for \eba, and 3300--3600 for \ebb.

We further note that the modeling of the spots is somewhat simplistic. We assume that there is only one large spot on the photosphere, as opposed to several small ones, and we assume that both stars are spotted (vs, e.g., one spotted and one unspotted). While the precise configuration of the modeled spots does not play a significant role in the derived spot sizes, contrast ratios, or all of the stellar parameters \citep{miller2021}, it is nonetheless one of the limitations of the model.

\subsection{Final Fitting}

We utilized emcee module within PHOEBE to solve the systems. In initial set up, we adopt \teff\ suggested by the APOGEE spectra \citep{sprague2022}, $q$ and $v_0$ for the systems derived in \citet{kounkel2021}, and the $P$ as measured by Lomb Scargle periodogram. $P$ required a very narrow initial distribution, as well as a prior, to ensure emcee would accurately converge, and not be pulled, e.g., towards the rotational period of a system. Other parameters were set up with reasonable distributions. The initial distribution guesses are included in Table \ref{tab:params}, and PHOEBE is able to improve on these distributions, including going beyond the bounds (e.g. for uniform distribution). An MCMC chain was set up with 250 walkers, and 25,000 steps; full convergence is achieved after $\sim$10,000 steps.

\begin{figure*}
\epsscale{1.2}
\plotone{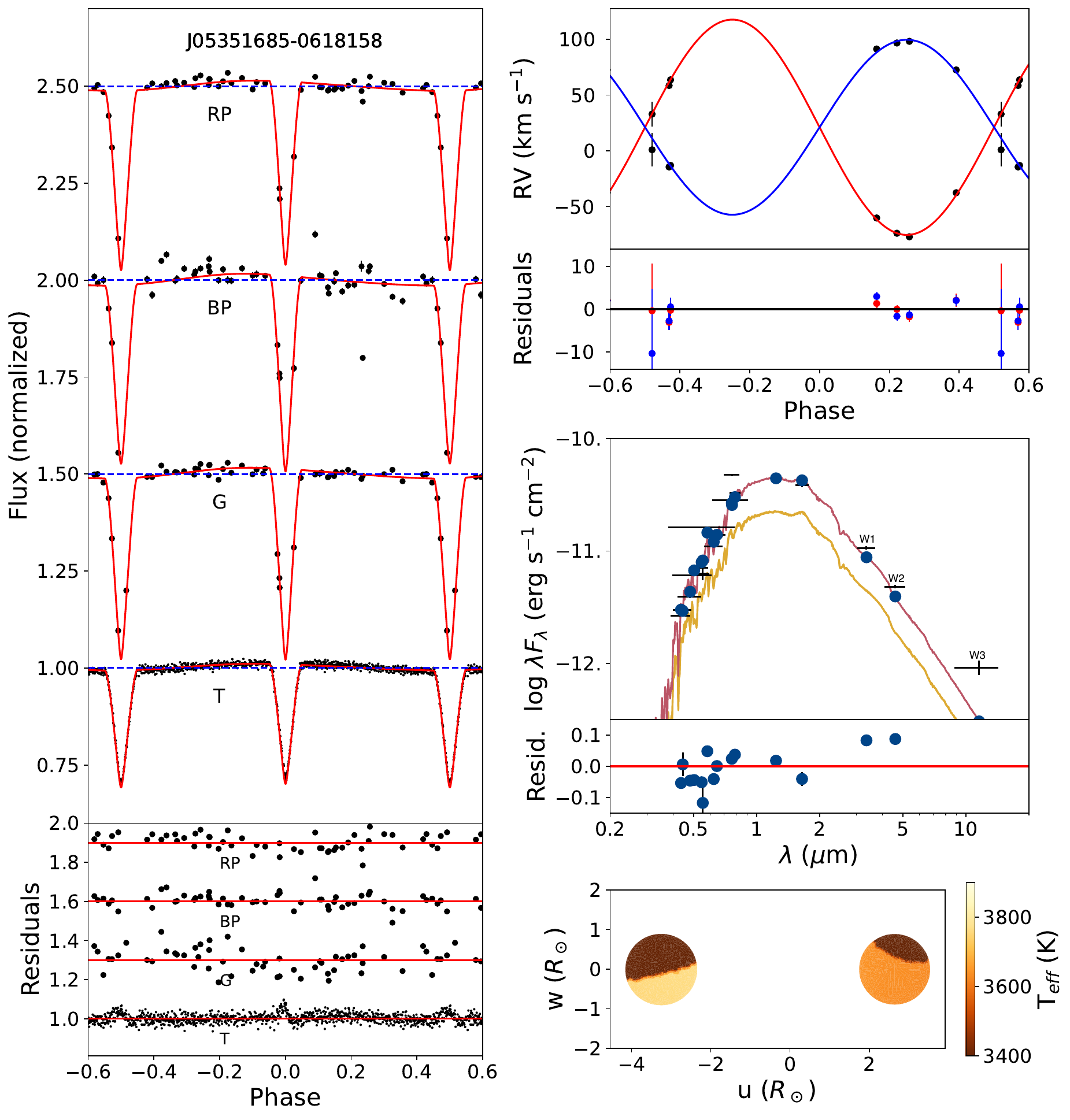}
\caption{Model fit of the observational data towards \eba. Left panel shows the fit of the available lightcurves (in  T, G, GP, and RP bandpasses) as well as the residuals (normalized by the typical uncertainties in the fluxes to meaningfully be displayed on the same plot). The blue line shows a flat continuum to demonstrate the significants of spots onto the light curve. Top right panel shows the fit of the radial velocities. Middle right panel shows the model SED (with the SED of the individual stars being shown in yellow, the combined SED in read, the black lines showing the measurements with the uncertainties and the bandpass width, and blue dots showing interpolated model fluxes onto these bandpasses). In this system, the last three fluxes, corresponding to W1, W2, and W3 passbands  appear to show IR excess. We also note that the two stars have extremely similar fluxes integrated across the entire surface, thus the two yellow lines are plotted on top of one another. Finally, the bottom right panel shows a mesh grid depicting the surface temperature, size, and the separation of the two stars.
\label{fig:eb1}}
\end{figure*}

\begin{figure*}
\epsscale{1.2}
\plotone{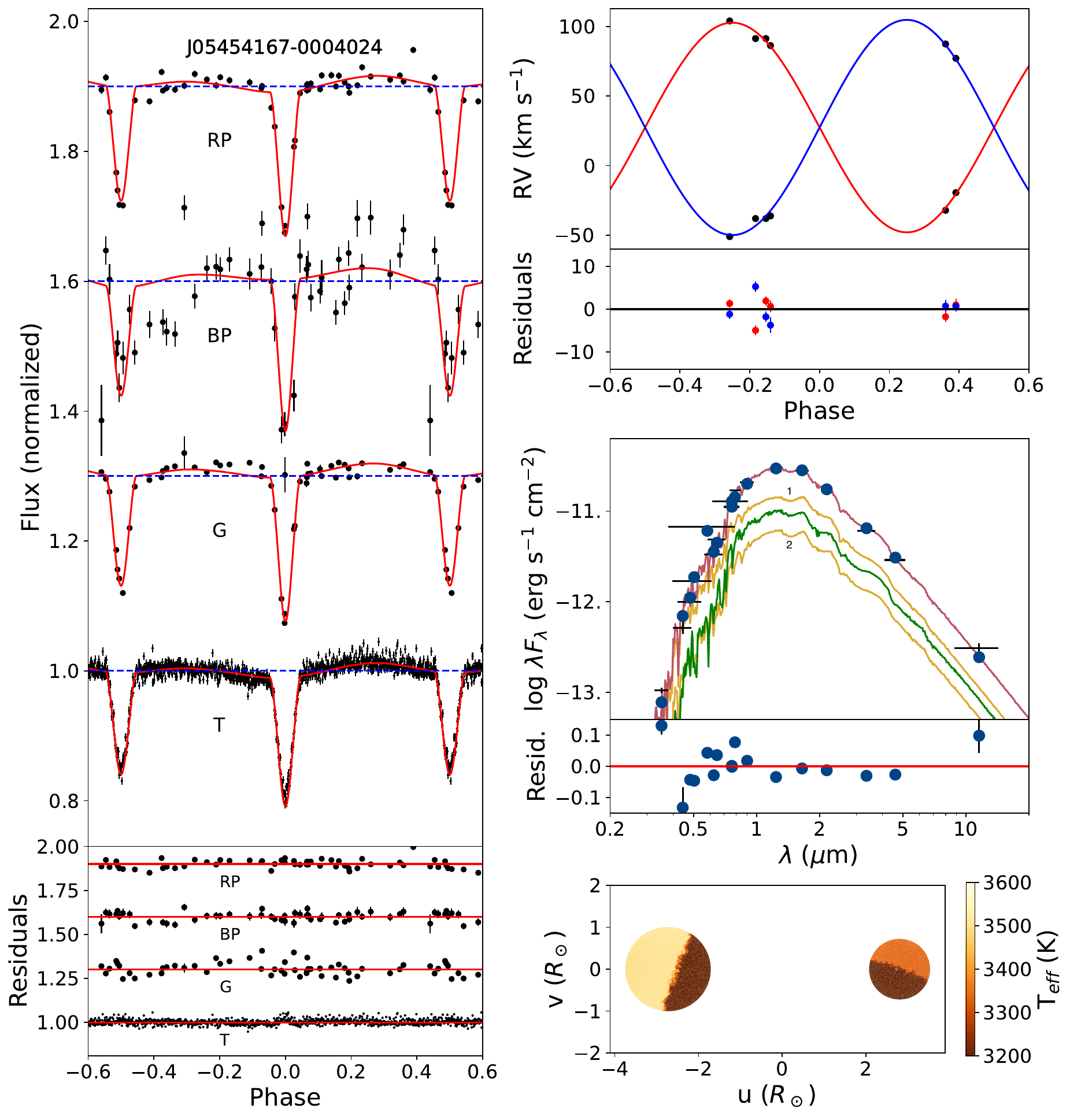}
\caption{Same as figure \ref{fig:eb1}, but for \ebb. In the middle right panel, SED marked with 1 corresponds to the primary in Table \ref{tab:params}, and 2 to the secondary. The green line is the fitted SED of the third star in the system that is needed in order to match the observed flux while keeping the system fixed at the reported distance from Gaia.
\label{fig:eb2}}
\end{figure*}

The derived parameters are listed in Table \ref{tab:params}. The resulting fit for the light curve and the RV curve are shown in Figures \ref{fig:eb1} and \ref{fig:eb2}.

\section{Results}\label{sec:result}

\begin{deluxetable*}{c|cc|cc}
\tablecaption{Stellar parameters of ESB2s
\label{tab:params}}
\tabletypesize{\scriptsize}
\tablewidth{\linewidth}
\tablehead{
  \colhead{Parameter} &
  \multicolumn{2}{|c}{\eba} &
  \multicolumn{2}{|c}{\ebb} \\
  \colhead{} &
  \multicolumn{1}{|c}{Initial} &
  \colhead{Fitted} &
  \multicolumn{1}{|c}{Initial} &
  \colhead{Fitted}
  }
\startdata
  $R_1$ (\rsun) &0.6--1.1& $0.902^{+0.029}_{-0.023}$ &0.6--1.1& $1.011^{+0.117}_{-0.093}$ \\
  $R_2$ (\rsun) &0.6--1.1& $0.891^{+0.022}_{-0.024}$ &0.6--1.1& $0.725^{+0.112}_{-0.077}$ \\
  $M_1$ (\msun) && $0.424^{+0.023}_{-0.023}$ && $0.348^{+0.018}_{-0.021}$ \\
  $M_2$ (\msun) && $0.522^{+0.029}_{-0.028}$ && $0.339^{+0.017}_{-0.019}$ \\
  $T_{\rm phot,1}$ (K) &3400--3500& $3753.0^{+47.0}_{-331.0}$ &3400$\pm$10& $3500.0^{+100.0}_{-200.0}$ \\
  $T_{\rm phot,2}$ (K) &3200--3600& $3630.0^{+54.0}_{-316.0}$ &3400$\pm$10& $3340.0^{+100.0}_{-190.0}$ \\
  $T_{\rm eff,1}$ (K) && $3483.0^{+46.0}_{-300.0}$ && $3313.0^{+153.0}_{-226.0}$ \\
  $T_{\rm eff,2}$ (K) && $3533.0^{+51.0}_{-300.0}$ && $3124.0^{+114.0}_{-194.0}$ \\
  $\log g_1$ && $4.155^{+0.026}_{-0.029}$ && $3.97^{+0.078}_{-0.091}$ \\
  $\log g_2$ && $4.256^{+0.025}_{-0.021}$ && $4.25^{+0.1}_{-0.11}$ \\
  \hline
  $P$ (d) &1.70439$\pm$1e-6& $1.704404^{+4.2e-07}_{-4.2e-07}$ &1.81719484$\pm$1e-6& $1.817212^{+1.8e-06}_{-1.7e-06}$ \\
  $a$ (\rsun) &5--7& $5.894^{+0.087}_{-0.09}$ &5--7& $5.532^{+0.085}_{-0.103}$ \\
  $i$ ($^\circ$) &85--90& $88.64^{+0.36}_{-0.28}$ &85--90& $82.3^{+1.6}_{-2.2}$ \\
  $q$ &1.2$\pm$0.05& $1.231^{+0.073}_{-0.067}$ &0.971$\pm$0.05& $0.974^{+0.041}_{-0.035}$ \\
  $t_0$ (MJD) &58986.407877$\pm$0.0001& $58986.40791^{+0.00013}_{-0.00016}$ &58986.6671513$\pm$0.0001& $58986.67893^{+0.00092}_{-0.00128}$ \\
  $\gamma$ (\kms) &20.9$\pm$1.5& $21.2^{+2.2}_{-2.0}$ &23.7$\pm$1& $27.4^{+1.2}_{-1.2}$ \\
  \hline
  $r_1$ ($^\circ$) &0--90& $89.75^{+0.25}_{-2.12}$ &0--90& $75.0^{+14.0}_{-17.0}$ \\
  $l_1$ ($^\circ$) &0--360& $255.8^{+7.8}_{-8.6}$ &0--360& $0.4^{+198.2}_{-7.1}$ \\
  $b_1$ ($^\circ$) &-90--90& $52.1^{+6.4}_{-9.3}$ &-90--90& $110.0^{+11.0}_{-43.0}$ \\
  $r_2$ ($^\circ$) &0--90& $52.1^{+6.4}_{-9.3}$&0--90& $83.7^{+6.2}_{-12.2}$ \\
  $l_2$ ($^\circ$) &0--360& $288.5^{+8.7}_{-8.0}$ &0--360& $4.9^{+45.8}_{-6.0}$ \\
  $b_2$ ($^\circ$) &-90--90& $286.0^{+2.4}_{-2.1}$ &-90--90& $160.0^{+28.0}_{-44.0}$ \\
  \hline
  $l3_{T}$ &0--0.5& $0.3212^{+0.0057}_{-0.0058}$ &0--0.8& $0.31^{+0.14}_{-0.22}$ \\
  $l3_{G}$ &0--0.05& $0.00055^{+0.00376}_{-0.00054}$ &0--0.8& $0.25^{+0.15}_{-0.24}$ \\
  $l3_{BP}$ &0--0.05& $0.005^{+0.0196}_{-0.005}$ &0--0.8& $0.24^{+0.19}_{-0.24}$ \\
  $l3_{RP}$ &0--0.05& $0.016^{+0.014}_{-0.014}$ &0--0.8& $0.24^{+0.15}_{-0.24}$ \\
\hline
 \av\ (mag) && 0.337 && 0.67 \\
 Dist. (pc) && 334 && 422 \\
\enddata
\end{deluxetable*}

\subsection{\eba}

Overall, the model for \eba\ is able to closely reproduce the observed SED of the system in the optical regime, although in WISE passbands there is a significant infrared excess due to the presence of a protoplanetary disk, although this system does not show evidence of accretion \citep{hsu2012,fang2013}.

The system consists of two stars with comparable luminosity and very similar radii of $\sim$0.9~\rsun. However, the masses of them are quite different, with the primary having a mass of $\sim0.52$~\msun, and the secondary having a mass of $\sim0.42$~\msun. Interestingly, the temperature of the photosphere in these two stars is backwards in comparison to the expectations, with the more massive star having cooler photosphere. This is explained by the less massive star having a more sizeable spot on its photosphere. PHOEBE estimates that the size of the spot on the secondary is $\sim90^\circ$, i.e, covering 50\% of the star -- this is the maximum spot size that is supported by PHOEBE, and the actual spot coverage may be larger, as spot coverage of $>50$\% is not uncommon in pre-main sequence stars \citep[e.g.,][]{gully-santiago2017}. The spot on the more massive star is smaller. Thus, calculating the true \teff\ does result in the more massive star in the system being hotter than the less massive star. 

Such a difference in spot sizes between the stars of different masses is consistent with pre-main sequence stars (in the \teff\ range of up to 4500 K) having stronger magnetic field strengths with increased mass, and thus having a larger discrepancy between temperatures of spotted and unspotted photospheres \citep{flores2022}.

\subsection{\ebb}

\ebb\ consists of two stars with near-equal masses of $\sim0.34$ \msun. However, they do have very different radii, one with the size of $\sim1.0$ \rsun, the other $0.7$ \rsun. The larger one appears to be hotter, and this is true both for the temperature of the photosphere as well as the true \teff, as both stars have comparable spot sizes.

Gaia DR3 parallax of \ebb\ is 2.3746$\pm$0.0493~mas, translating to a likely distance range of 412.6--430.1~pc. SED fitting using the derived stellar parameters at this distance results in a significant underestimation of the total observed flux towards this system across most of the electromagnetic spectrum.

A better fit can be achieved if the distance is adjusted to 360~pc, however, PHOEBE fit suggests a third light of $\sim$0.31 for TESS passband, and $\sim0.25$ for Gaia passpands. Since Gaia does have very high spatial resolution, it is highly improbable for the third light contamination to originate from a chance alignment of an unrelated star along the same line of sight. Rather, it is likely that \ebb\ is a triple system, with a tertiary having a significantly longer orbital period. Indeed, allowing inclusion of a third object while doing an SED fit (at the distance reported by Gaia) favors a presence of a star with \teff$\sim$3130~K and $R\sim$0.92~\rsun. This is a similar \teff\ as a more compact star in the system, and the estimated radius is in between that of either star in the inner binary. Such a star typically has luminosity which is 20--30\% that of the combined luminosity of the inner binary, consistent with the estimate for the $l3$ that is seen across all passbands.

\subsection{Degeneracy in the radii}

In eclipsing binaries with circular orbits, the radius parameter that is most stringently constrained is the sum of the radii. Although extreme combinations of the radii can be ruled out based on the bottom of the light curve during the eclipse, it is possible to redistribute the radii of the individual stars within a certain range of the sum and still produce a comparable light curve fit. Thus, external confirmation is often useful.

Examining the cross-correlation function of the spectrum consisting of two stars of comparable temperature can be used for such a purpose. The ratio of area under the peak of the CCF is roughly proportional to the ratio of luminosities at the spectral bandpass. We note that this is an approximation that does have its limitations - e.g. in stars of very different temperatures affecting the strength of the cross-correlation depending on the template that is being used \citep{kounkel2024}.

In a heavily spotted system, such as the ones in this work, this approximation has further complications. The ratio of the luminosities of two stars can vary depending on the phase, depending on the side of a given star that is facing us. Indeed, the ratio of the area under the CCF peaks does vary at different epochs. Unfortunately, while we are able to create a model of spots based on the lightcurves, this model is fairly simplistic, and given that the morphology and the position of spots can vary significantly in time, even if the simplistic model was completely accurate, it may not necessarily hold at the time the spectra was taken, as it was several years prior to the TESS observations. Nonetheless, from the rough estimate of the ratio of the luminosities, we can approximate the ratio of the radii.

The typical ratio of the area under the CCF peaks between the secondary and the primary for is 1.2--1.4 for \eba, and 0.5--0.7 for \ebb. Using the respective \teff\ for all of the stars, this results in the estimate of the ratio of the radii of 1.0--1.1 for \eba, and 0.8--0.9 for \ebb. Combining these ratios with the sum of the radii derived by PHOEBE produces the radii of 0.87+0.91 \rsun\ for \eba, and 0.91+0.82 \rsun\ for \ebb, which is consistent with the reported uncertainties in Table \ref{tab:params}.

\section{Discussion}\label{sec:discussion}

\subsection{Ages}

\begin{figure*}
\epsscale{1.1}
\plottwo{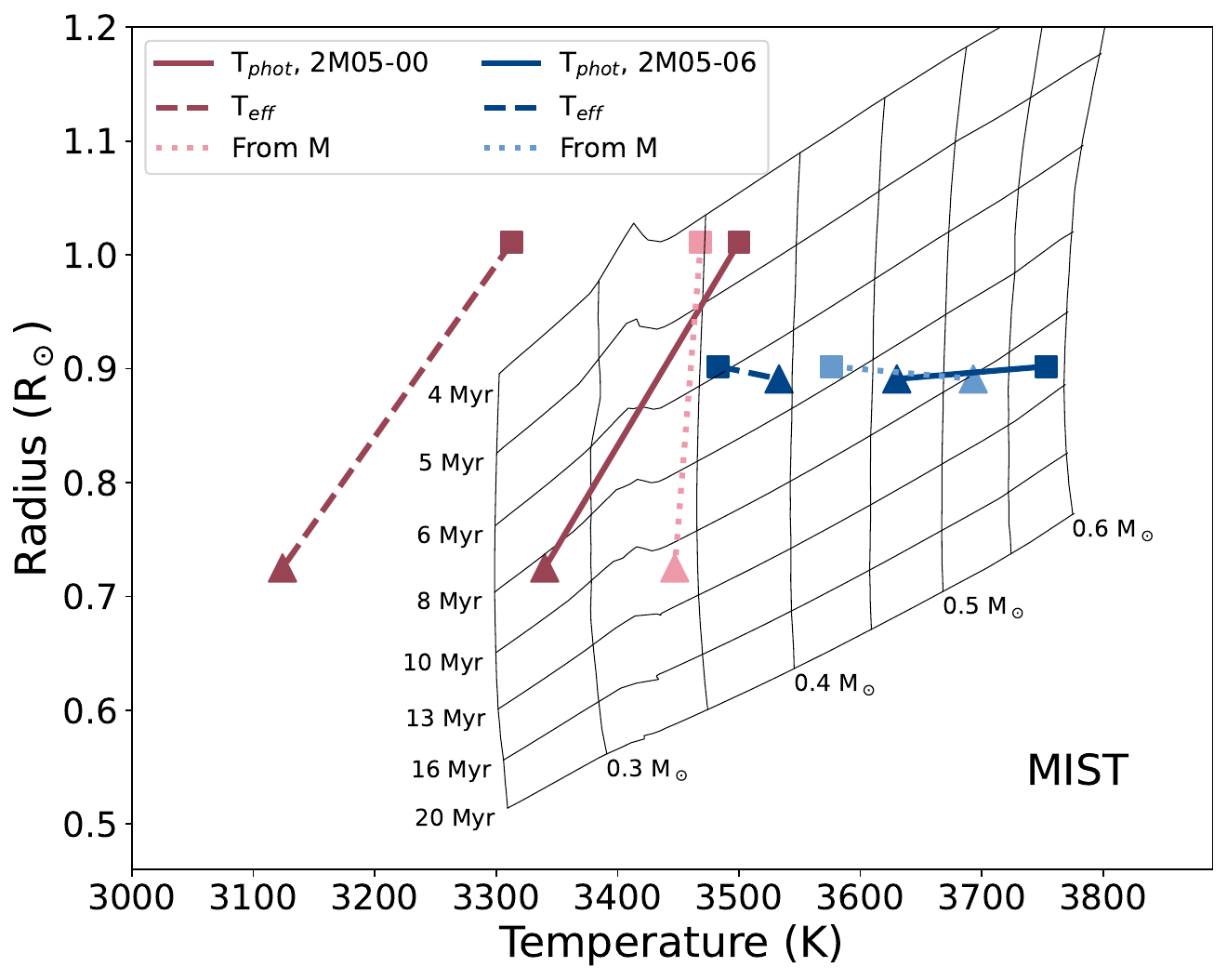}{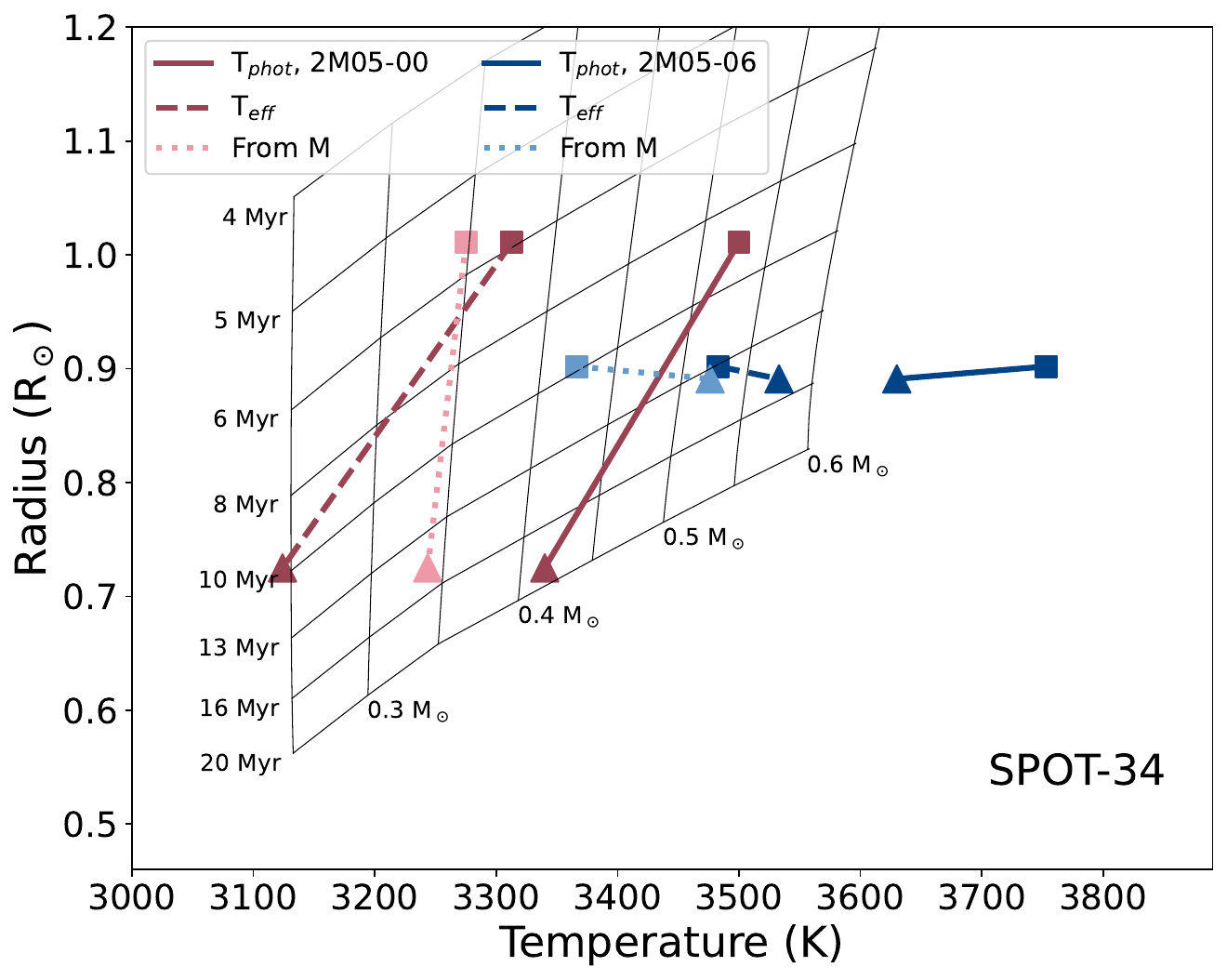}
\plottwo{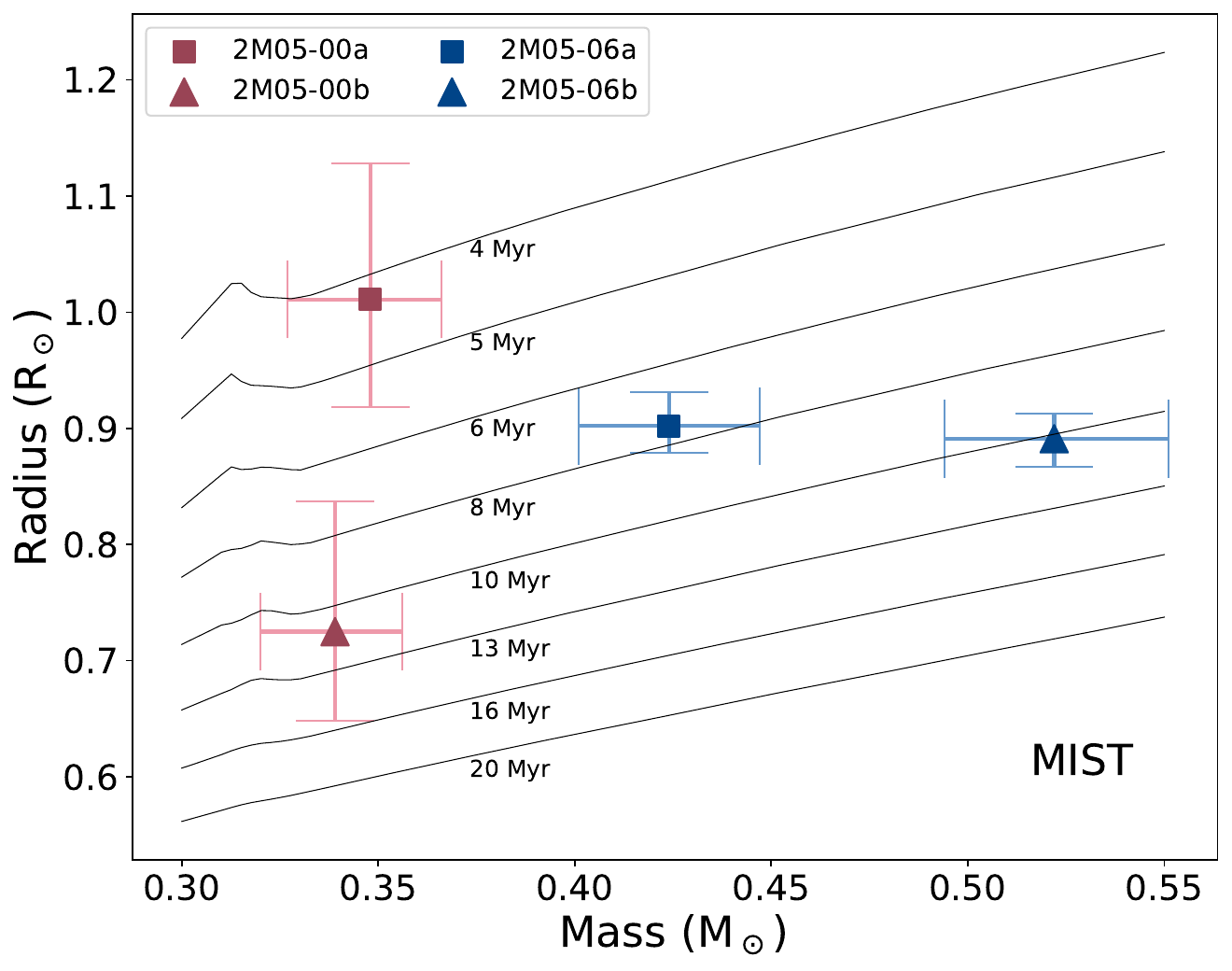}{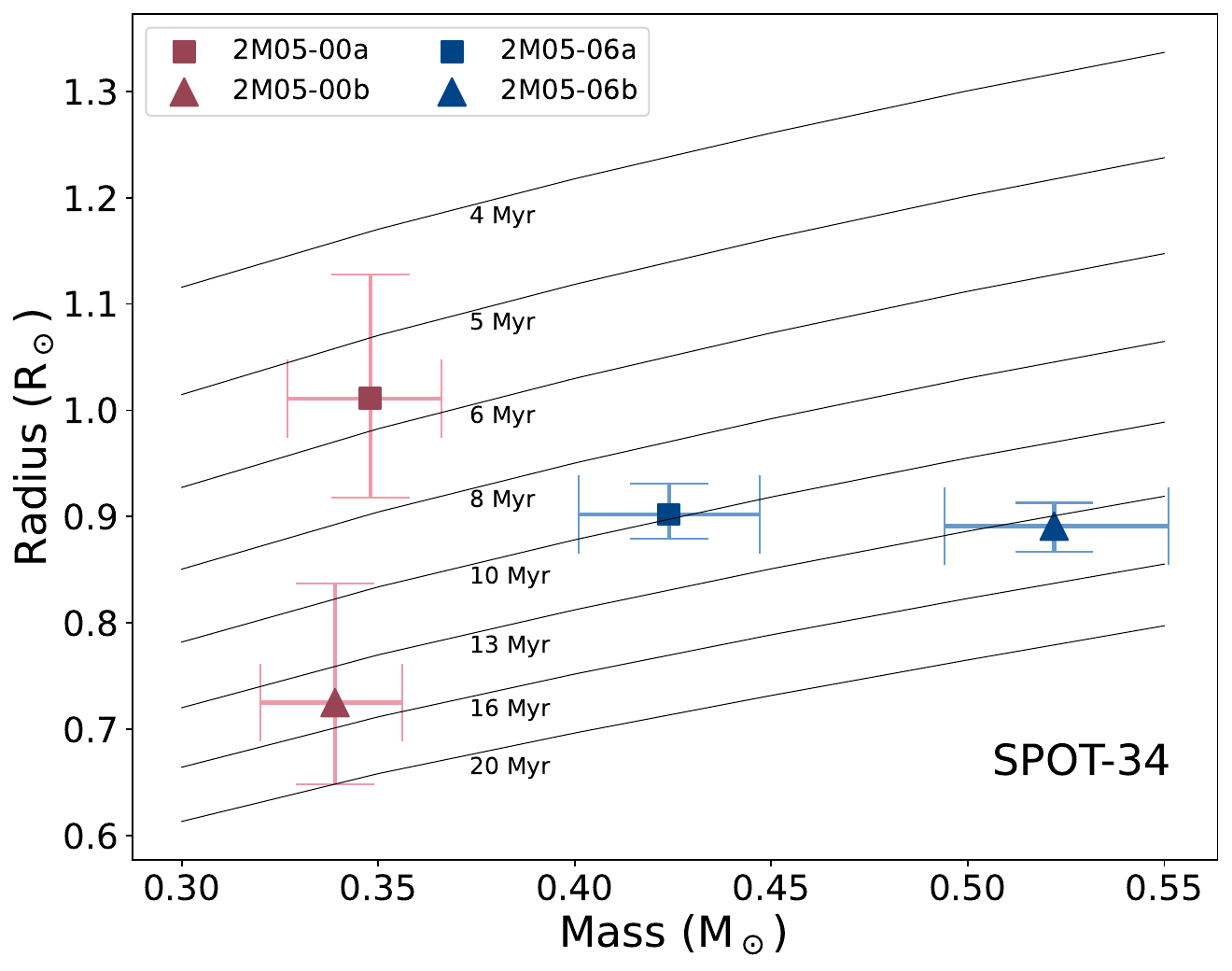}
\plottwo{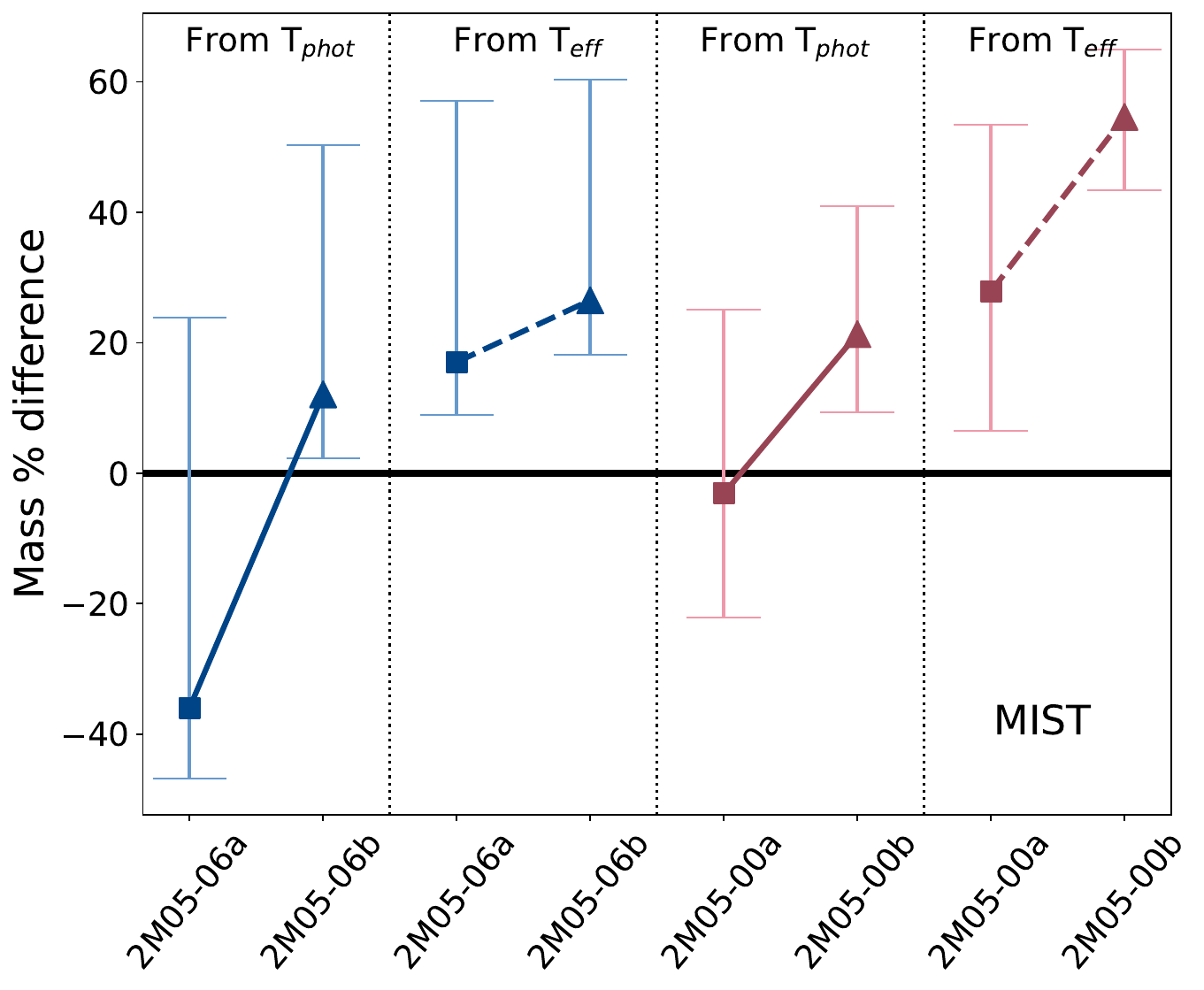}{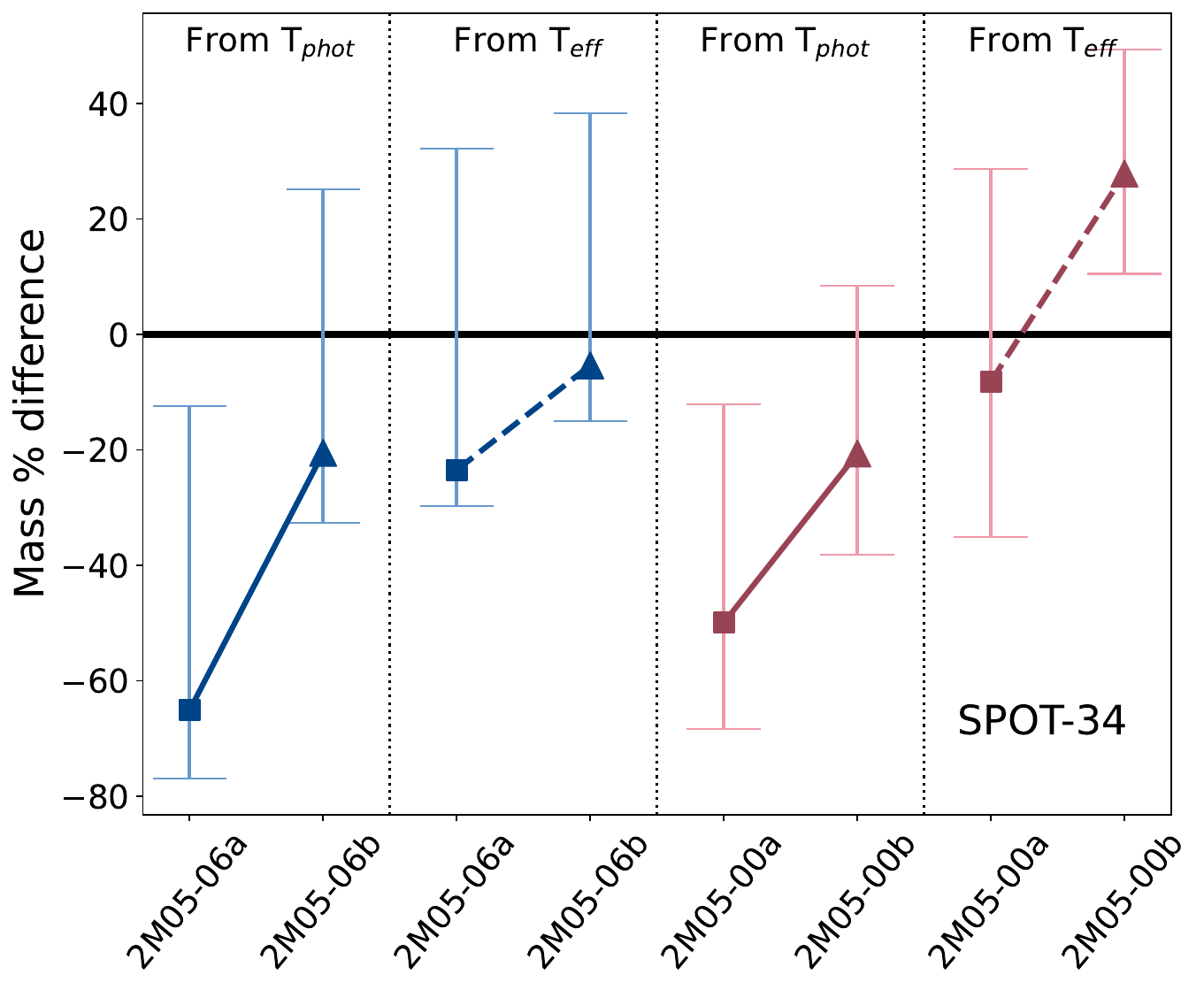}
\caption{Top: Effective temperature (dashed line) and the temperature of the photosphere (solid line) for \ebb\ (red) and \eba\ (blue) versus their radii. The transformation of this parameter space onto mass and age using isochrones is also shown, alongside the masses of the stars in these two systems as a function of radius. Middle: Mass vs radius relative to the isochrones, showing the typical uncertainty in the measurements. Bottom: the difference between the observed masses (from the orbital fitting) and the interpolated model mass from temperatures, divided by the observed mass. Error bars include the uncertainty in the temperature. Left: non-magnetic MIST isochrones \citep{choi2016}. Right: magnetic SPOT isochrones \citep{somers2020}, with 34\% spot filling fraction.
\label{fig:iso}}
\end{figure*}

Most commonly used method to determine ages of pre-main sequence stars is through interpolating their photometry over the isochrones, but ages of the binary stars are commonly underestimated since they have higher total luminosity in comparison to single stars. However, since we have determined all of the fundamental parameters for these stars, we can use them to estimate ages of the individual stars in the system instead. We consider two projections: mass vs radius, as well as \teff\ (and $T_{\rm phot}$) vs radius (Figure \ref{fig:iso}. We compare our observations to MIST \citep{choi2016} isochrones; although we considered other models, they do not have a significant impact on the results.

\ebb\ appears to be a system of two stars that appear to be mostly coeval, with the age of 6.9--7.1~dex, or 8--12~Myr depending on the precise projection that is used for evaluation. This is broadly consistent with the system being in the foreground of Orion A molecular cloud and being a member of much more evolved Orion D \citep{kounkel2018a}.

However, \eba\ appears to be somewhat peculiar. In this system, the primary favors the age of 6.7~dex ($\sim$5~Myr), while the secondary favors the age of 7~dex ($\sim$10~Myr). Such a large apparent difference in age is highly unusual, as typically young EBs do tend to show coevality to within $\sim$20\% \citep{stassun2014}, and indeed this is the expectation from standard binary star formation theory.

Given that \eba\ is a triple star system, it is not impossible to consider that the tertiary companion may have been sufficiently close as to the inner binary so that its orbital motion would have affected the observed ratio of the radial velocities over the 245 day baseline over which RVs have been obtained. This could have had an impact on the observed mass ratio, making the determination of the masses to be incorrect.

This, however, would not propagate to \teff s, as \teff\ ratio is determined purely from the light curves (with the absolute calibration assisted by the SED). Similarly, while the absolute calibration of the individual radii does depend on the orbital speed, the ratio of the radii is driven solely by the eclipse geometry. The apparent factor of 2 difference between the ages of the stars persists in \teff\ vs radius projection.

\eba\ is found in the vicinity of NGC 2068 cluster in Orion B molecular cloud which has a typical age of 1--2 Myr \citep{kounkel2018a}, and although in Orion as a whole there are a number of older stars, there are very few of them near NGC 2068. Thus, of the two derived ages, younger age is significantly favored for this system, and a chance dynamical capture of an isolated older star is highly unlikely. It is unclear what type of activity could be responsible for such a large disparity.

\subsection{\teff}

Surface temperature is strongly correlated with mass. As pre--main-sequence stars evolve down the Hayashi track, there is almost no dependence on age between mass and temperature, resulting in a robust relation.

Spots, however, introduce a significant complication. Masses derived dynamically, in comparison to the masses derived from the temperature, can have 50--100\% discrepancy \citep[e.g.,][]{david2019,flores2022}. We observe this here also. \teff s that have been derived for the systems in these work are generally consistent with spectroscopic estimates \citep{fang2009,hsu2012,sprague2022}, pointing to their accuracy. These \teff s significantly underestimate the masses we derive from the full orbital fit by 20--50\% when using non-magnetic isochrones such as MIST \citep{choi2016}, and the difference is especially pronounced in the case of \eba. Masses derived $T_{\rm phot}$ are a much closer fit to the dynamical masses. On the other hand, comparison made using isochrones that do consider magnetic activity, such as SPOTS tracks \citep{somers2020} with 34\% spot coverage do produce a much better agreement between the dynamical masses and the observed \teff s. This set of models produced the best agreement between these two parameters; we also examined isochrones with 17\% and 51\%, with the former underestimating the mass, and the latter overestimating the mass.

\subsection{Comparison to other systems}

While the eclipsing binaries shown in this paper do present some peculiarities, they are not unique. For example, \ebb\ has a remarkable similarity to Parenago 1802, a $\sim$1 Myr system in Orion consisting of two stars of identical masses, but with radii discrepancy of $\sim$7\%, \teff\ discrepancy of $\sim$9\%, which results in a luminosity discrepancy of $\sim$60\%, which also has an appearance of not being coeval \citep{stassun2008}. Similarly to \ebb, it appears to be a triple system, and it has been interpreted that the tidal interactions could be responsible for the difference in the apparent luminosity of the binary \citep{gomez-maqueo-chew2012}.

On the other hand, \eba\ can be compared to 2MASS J05352184-0546085, a system of eclipsing brown dwarfs in Orion, where the lower mass star in the system appears to be hotter than the primary \citep{stassun2007}, due to the primary having stronger chromospheric activity than the secondary, suppressing its temperature and inflating its radius \citep{stassun2012}.

\section{Conclusions}\label{sec:concl}

In this work we present the analysis of the two young eclipsing binaries \eba\ and \ebb. With ages $<$10 Myr, they are valuable benchmarks of testing stellar evolutionary models, as the number of such systems still counts in a few dozen \citep[e.g.][]{stassun2014,david2019}.
These two systems, however, present interesting properties that differ from the predictions of standard stellar models and that therefore provide insight into the physical mechanisms that can alter the fundamental stellar properties of young stars. 

One of them, \eba, appears to be a mostly coeval system with two stars of different mass ($\sim$0.52 and $\sim$0.42~\msun), but very similar radius ($\sim$0.9~\rsun). With the less massive star being very heavily spotted, it actually has a hotter temperature of the (unspotted) photosphere in comparison to the more massive star.

The other, \ebb\ (a likely hierarchical triple system), has the inner binary with very similar masses($\sim$0.34~\msun), but very different radii ($\sim$0.7 and $\sim$1.0~\rsun). This creates the appearance that these stars are not coeval, one with the age of $\sim$4~Myr and the other with the age of $\sim$10~Myr according to standard stellar isochrones, with the population in which they are found favoring the younger age. 
%It is unclear what physical processes or activity could have led to such an apparent difference in age.
Similar behavior has been reported previously for at least one other young EB \citep[Par~1802; see][]{stassun2008,gomez-maqueo-chew2012}, which similarly possesses a tertiary whose dynamical heating has been suggested to possibly explain the anomalous properties of the inner binary \citep{gomez-maqueo-chew2019}. 

We also find that spots play a significant role in affecting the overall temperature of a star, and that effective temperature appears to significantly underestimate dynamical masses in these EBs when considering non-magnetic isochrones; magnetic tracks offer a much better agreement. Alternatively, the temperature of the unspotted photosphere appears to be in agreement with the non-magnetic isochrones. 

As directly measured masses of more pre--main-sequence stars are measured, the improved temperature determination of both spots and photospheres \citep[such as in EBs as in this work, or through spectral decomposition, e.g.,][]{gully-santiago2017}
%a better relationship between masses and temperatures could be established.
will permit a fuller assessment of the role of magnetism in altering the fundamental physical properties of young stars.

\section*{Acknowledgments}
MK acknowledges support provided by NASA grant 80NSSC24K0620.

This work has made use of data from the European Space Agency (ESA) mission {\it Gaia} (\url{https://www.cosmos.esa.int/gaia}), processed by the {\it Gaia} Data Processing and Analysis Consortium (DPAC, \url{https://www.cosmos.esa.int/web/gaia/dpac/consortium}). Funding for the DPAC has been provided by national institutions, in particular the institutions participating in the {\it Gaia} Multilateral Agreement.

%\begin{figure*}
%\epsscale{1.0}
%		\gridline{\fig{fig1.pdf}{0.25\textwidth}{}
%		          \fig{fig2.pdf}{0.25\textwidth}{}
%        }\vspace{-1cm}
%\caption{Caption
%\label{fig:figure}}
%\end{figure*}

\software{PHOEBE \citep{conroy2020}}

\bibliographystyle{aasjournal.bst}
%\bibliography{references.bib}
\bibliography{main.bbl}

\begin{thebibliography}{}
\expandafter\ifx\csname natexlab\endcsname\relax\def\natexlab#1{#1}\fi
\providecommand{\url}[1]{\href{#1}{#1}}
\providecommand{\dodoi}[1]{doi:~\href{http://doi.org/#1}{\nolinkurl{#1}}}
\providecommand{\doeprint}[1]{\href{http://ascl.net/#1}{\nolinkurl{http://ascl.net/#1}}}
\providecommand{\doarXiv}[1]{\href{https://arxiv.org/abs/#1}{\nolinkurl{https://arxiv.org/abs/#1}}}

\bibitem[{{Allard} {et~al.}(2011){Allard}, {Homeier}, \&
  {Freytag}}]{allard2011}
{Allard}, F., {Homeier}, D., \& {Freytag}, B. 2011, in Astronomical Society of
  the Pacific Conference Series, Vol. 448, 16th Cambridge Workshop on Cool
  Stars, Stellar Systems, and the Sun, ed. C.~{Johns-Krull}, M.~K. {Browning},
  \& A.~A. {West}, 91, \dodoi{10.48550/arXiv.1011.5405}

\bibitem[{{Berdyugina}(2005)}]{berdyugina2005}
{Berdyugina}, S.~V. 2005, Living Reviews in Solar Physics, 2, 8,
  \dodoi{10.12942/lrsp-2005-8}

\bibitem[{{Choi} {et~al.}(2016){Choi}, {Dotter}, {Conroy}, {Cantiello},
  {Paxton}, \& {Johnson}}]{choi2016}
{Choi}, J., {Dotter}, A., {Conroy}, C., {et~al.} 2016, \apj, 823, 102,
  \dodoi{10.3847/0004-637X/823/2/102}

\bibitem[{{Conroy} {et~al.}(2020){Conroy}, {Kochoska}, {Hey}, {Pablo},
  {Hambleton}, {Jones}, {Giammarco}, {Abdul-Masih}, \&
  {Pr{\v{s}}a}}]{conroy2020}
{Conroy}, K.~E., {Kochoska}, A., {Hey}, D., {et~al.} 2020, \apjs, 250, 34,
  \dodoi{10.3847/1538-4365/abb4e2}

\bibitem[{{David} {et~al.}(2019){David}, {Hillenbrand}, {Gillen}, {Cody},
  {Howell}, {Isaacson}, \& {Livingston}}]{david2019}
{David}, T.~J., {Hillenbrand}, L.~A., {Gillen}, E., {et~al.} 2019, \apj, 872,
  161, \dodoi{10.3847/1538-4357/aafe09}

\bibitem[{{Fang} {et~al.}(2013){Fang}, {Kim}, {van Boekel}, {Sicilia-Aguilar},
  {Henning}, \& {Flaherty}}]{fang2013}
{Fang}, M., {Kim}, J.~S., {van Boekel}, R., {et~al.} 2013, \apjs, 207, 5,
  \dodoi{10.1088/0067-0049/207/1/5}

\bibitem[{{Fang} {et~al.}(2009){Fang}, {van Boekel}, {Wang}, {Carmona},
  {Sicilia-Aguilar}, \& {Henning}}]{fang2009}
{Fang}, M., {van Boekel}, R., {Wang}, W., {et~al.} 2009, \aap, 504, 461,
  \dodoi{10.1051/0004-6361/200912468}

\bibitem[{{Feinstein} {et~al.}(2019){Feinstein}, {Montet}, {Foreman-Mackey},
  {Bedell}, {Saunders}, {Bean}, {Christiansen}, {Hedges}, {Luger}, {Scolnic},
  \& {Cardoso}}]{feinstein2019}
{Feinstein}, A.~D., {Montet}, B.~T., {Foreman-Mackey}, D., {et~al.} 2019,
  \pasp, 131, 094502, \dodoi{10.1088/1538-3873/ab291c}

\bibitem[{{Flores} {et~al.}(2022){Flores}, {Connelley}, {Reipurth}, \&
  {Duch{\^e}ne}}]{flores2022}
{Flores}, C., {Connelley}, M.~S., {Reipurth}, B., \& {Duch{\^e}ne}, G. 2022,
  \apj, 925, 21, \dodoi{10.3847/1538-4357/ac37bd}

\bibitem[{{Gaia Collaboration} {et~al.}(2022){Gaia Collaboration}, {Vallenari},
  {Brown}, {Prusti}, {de Bruijne}, {Arenou}, {Babusiaux}, {Biermann},
  {Creevey}, {Ducourant}, \& et~al.}]{gaia-collaboration2022a}
{Gaia Collaboration}, {Vallenari}, A., {Brown}, A.~G.~A., {et~al.} 2022, arXiv
  e-prints, arXiv:2208.00211.
\newblock \doarXiv{2208.00211}

\bibitem[{{G{\'o}mez Maqueo Chew} {et~al.}(2012){G{\'o}mez Maqueo Chew},
  {Stassun}, {Pr{\v{s}}a}, {Stempels}, {Hebb}, {Barnes}, {Heller}, \&
  {Mathieu}}]{gomez-maqueo-chew2012}
{G{\'o}mez Maqueo Chew}, Y., {Stassun}, K.~G., {Pr{\v{s}}a}, A., {et~al.} 2012,
  \apj, 745, 58, \dodoi{10.1088/0004-637X/745/1/58}

\bibitem[{{G{\'o}mez Maqueo Chew} {et~al.}(2019){G{\'o}mez Maqueo Chew},
  {Hebb}, {Stempels}, {Paat}, {Stassun}, {Faedi}, {Street}, {Rohn}, {Hellier},
  \& {Anderson}}]{gomez-maqueo-chew2019}
{G{\'o}mez Maqueo Chew}, Y., {Hebb}, L., {Stempels}, H.~C., {et~al.} 2019,
  \aap, 623, A23, \dodoi{10.1051/0004-6361/201833299}

\bibitem[{{Gully-Santiago} {et~al.}(2017){Gully-Santiago}, {Herczeg},
  {Czekala}, {Somers}, {Grankin}, {Covey}, {Donati}, {Alencar}, {Hussain},
  {Shappee}, {Mace}, {Lee}, {Holoien}, {Jose}, \& {Liu}}]{gully-santiago2017}
{Gully-Santiago}, M.~A., {Herczeg}, G.~J., {Czekala}, I., {et~al.} 2017, \apj,
  836, 200, \dodoi{10.3847/1538-4357/836/2/200}

\bibitem[{{Hsu} {et~al.}(2012){Hsu}, {Hartmann}, {Allen}, {Hern{\'a}ndez},
  {Megeath}, {Mosby}, {Tobin}, \& {Espaillat}}]{hsu2012}
{Hsu}, W.-H., {Hartmann}, L., {Allen}, L., {et~al.} 2012, \apj, 752, 59,
  \dodoi{10.1088/0004-637X/752/1/59}

\bibitem[{{Kounkel}(2023)}]{sedfit}
{Kounkel}, M. 2023, mkounkel/SEDFit: 0.3, 0.3,  Zenodo,
  \dodoi{10.5281/zenodo.8076501}

\bibitem[{{Kounkel} {et~al.}(2024){Kounkel}, {Statti}, {Kulkarni}, {Stassun},
  \& {Sun}}]{kounkel2024}
{Kounkel}, M., {Statti}, M., {Kulkarni}, A., {Stassun}, K.~G., \& {Sun}, M.
  2024, \mnras, 527, 3806, \dodoi{10.1093/mnras/stad3439}

\bibitem[{{Kounkel} {et~al.}(2018){Kounkel}, {Covey}, {Su{\'a}rez},
  {Rom{\'a}n-Z{\'u}{\~n}iga}, {Hernandez}, {Stassun}, {Jaehnig}, {Feigelson},
  {Pe{\~n}a Ram{\'\i}rez}, {Roman-Lopes}, {Da Rio}, {Stringfellow}, {Kim},
  {Borissova}, {Fern{\'a}ndez-Trincado}, {Burgasser},
  {Garc{\'\i}a-Hern{\'a}ndez}, {Zamora}, {Pan}, \& {Nitschelm}}]{kounkel2018a}
{Kounkel}, M., {Covey}, K., {Su{\'a}rez}, G., {et~al.} 2018, \aj, 156, 84,
  \dodoi{10.3847/1538-3881/aad1f1}

\bibitem[{{Kounkel} {et~al.}(2021){Kounkel}, {Covey}, {Stassun},
  {Price-Whelan}, {Holtzman}, {Chojnowski}, {Longa-Pe{\~n}a},
  {Rom{\'a}n-Z{\'u}{\~n}iga}, {Hernandez}, {Serna}, {Badenes}, {De Lee},
  {Majewski}, {Stringfellow}, {Kratter}, {Moe}, {Frinchaboy}, {Beaton},
  {Fern{\'a}ndez-Trincado}, {Mahadevan}, {Minniti}, {Beers}, {Schneider},
  {Barba}, {Brownstein}, {Garc{\'\i}a-Hern{\'a}ndez}, {Pan}, \&
  {Bizyaev}}]{kounkel2021}
{Kounkel}, M., {Covey}, K.~R., {Stassun}, K.~G., {et~al.} 2021, \aj, 162, 184,
  \dodoi{10.3847/1538-3881/ac1798}

\bibitem[{{Miller} {et~al.}(2021){Miller}, {Kounkel}, {Boggio}, {Covey}, \&
  {Price-Whelan}}]{miller2021}
{Miller}, A., {Kounkel}, M., {Boggio}, C., {Covey}, K., \& {Price-Whelan},
  A.~M. 2021, \pasp, 133, 044201, \dodoi{10.1088/1538-3873/abeaf7}

\bibitem[{{Pecaut} \& {Mamajek}(2013)}]{pecaut2013}
{Pecaut}, M.~J., \& {Mamajek}, E.~E. 2013, \apjs, 208, 9,
  \dodoi{10.1088/0067-0049/208/1/9}

\bibitem[{{Pr{\v{s}}a} {et~al.}(2008){Pr{\v{s}}a}, {Guinan}, {Devinney},
  {DeGeorge}, {Bradstreet}, {Giammarco}, {Alcock}, \& {Engle}}]{prsa2008}
{Pr{\v{s}}a}, A., {Guinan}, E.~F., {Devinney}, E.~J., {et~al.} 2008, \apj, 687,
  542, \dodoi{10.1086/591783}

\bibitem[{{Somers} {et~al.}(2020){Somers}, {Cao}, \&
  {Pinsonneault}}]{somers2020}
{Somers}, G., {Cao}, L., \& {Pinsonneault}, M.~H. 2020, \apj, 891, 29,
  \dodoi{10.3847/1538-4357/ab722e}

\bibitem[{{Sprague} {et~al.}(2022){Sprague}, {Culhane}, {Kounkel}, {Olney},
  {Covey}, {Hutchinson}, {Lingg}, {Stassun}, {Rom{\'a}n-Z{\'u}{\~n}iga},
  {Roman-Lopes}, {Nidever}, {Beaton}, {Borissova}, {Stutz}, {Stringfellow},
  {Ram{\'\i}rez}, {Ram{\'\i}rez-Preciado}, {Hern{\'a}ndez}, {Kim}, \&
  {Lane}}]{sprague2022}
{Sprague}, D., {Culhane}, C., {Kounkel}, M., {et~al.} 2022, \aj, 163, 152,
  \dodoi{10.3847/1538-3881/ac4de7}

\bibitem[{{Stassun} {et~al.}(2014){Stassun}, {Feiden}, \&
  {Torres}}]{stassun2014}
{Stassun}, K.~G., {Feiden}, G.~A., \& {Torres}, G. 2014, \nar, 60, 1,
  \dodoi{10.1016/j.newar.2014.06.001}

\bibitem[{{Stassun} {et~al.}(2012){Stassun}, {Kratter}, {Scholz}, \&
  {Dupuy}}]{stassun2012}
{Stassun}, K.~G., {Kratter}, K.~M., {Scholz}, A., \& {Dupuy}, T.~J. 2012, \apj,
  756, 47, \dodoi{10.1088/0004-637X/756/1/47}

\bibitem[{{Stassun} {et~al.}(2008){Stassun}, {Mathieu}, {Cargile}, {Aarnio},
  {Stempels}, \& {Geller}}]{stassun2008}
{Stassun}, K.~G., {Mathieu}, R.~D., {Cargile}, P.~A., {et~al.} 2008, \nat, 453,
  1079, \dodoi{10.1038/nature07069}

\bibitem[{{Stassun} {et~al.}(2007){Stassun}, {Mathieu}, \&
  {Valenti}}]{stassun2007}
{Stassun}, K.~G., {Mathieu}, R.~D., \& {Valenti}, J.~A. 2007, \apj, 664, 1154,
  \dodoi{10.1086/519231}

\end{thebibliography}

\end{document}